\documentclass[12pt]{article}
\usepackage{epsfig}

\usepackage{amssymb}
\usepackage{amsmath}
\usepackage{amsfonts}
\usepackage{amsthm}

\theoremstyle{theorem}

\theoremstyle{definition}

\def\bp{\begin{proof}}
\def\ep{\end{proof}}

\def\be{\begin{equation}}
\def\ee{\end{equation}}
\def\ba{\begin{array}{c}}
\def\ea{\end{array}}

\def\ben{$$}
\def\een{$$}

\newcommand{\bea}{\begin{eqnarray}}
\newcommand{\eea}{\end{eqnarray}}

\newcommand{\kt}{\rangle}
\newcommand{\br}{\langle}
\begin{document}

\titlepage

\vspace{.35cm}

 \begin{center}{\Large \bf

Decays of degeneracies in  ${\cal PT}-$symmetric ring-shaped
 lattices

  }\end{center}

\vspace{10mm}

 \begin{center}

 {\bf Miloslav Znojil}

 \vspace{3mm}
Nuclear Physics Institute ASCR,

250 68 \v{R}e\v{z}, Czech Republic

{e-mail: znojil@ujf.cas.cz}

\vspace{3mm}


\end{center}

\vspace{5mm}


\section*{Abstract}

Many apparently non-Hermitian ring-shaped discrete lattices share
the appeal with their more popular linear predecessors.
Irrespectively of the change of the topology, their dynamics
controlled by the nearest-neighbor interaction is shown to remain
phenomenologically interesting. For the perturbation-caused removals
of spectral degeneracy at  exceptional points, in particular,
alternative scenarios are illustrated via solvable examples.

\newpage

\section{Introduction}

%
%

The concept of ${\cal PT}-$symmetry \cite{Carl} is an inspiring
source of experimental activities in several branches of physics
like optics \cite{Makris} or solid-state physics
\cite{Joglekar,Joglekarbe} or magnetohydrodynamics \cite{Eva}. The
idea itself originates in quantum mechanics. In a way illustrated by
numerous toy models \cite{BB,Geza} the interest in ${\cal
PT}-$symmetry has been motivated there by a counterintuitive
contrast between the manifestly non-Hermitian form of a given ${\cal
PT}-$symmetric interaction Hamiltonian $H$ and the reality of the
spectrum of the related energies inside a certain domain ${\cal D}$
of optional parameters.

The mathematical key to the puzzling existence of a non-empty domain
${\cal D}$  of parameters giving real spectra has been found in
Ref.~\cite{Geyer}. The authors emphasized there the compatibility of
the non-Hermiticity of $H $ in a given Hilbert space (say, in ${\cal
H}^{(friendly)}:=L^2(\mathbb{R})$ where $H \neq H^\dagger$) with the
{\em simultaneous} Hermiticity of the same operator in another
Hilbert space. In particular, the new space ${\cal
H}^{(sophisticated)}$ may be allowed to coincide with ${\cal
H}^{(friendly)}$ as a vector space, being just endowed with another,
nontrivial, {\em ad hoc} inner product.

The main phenomenological appeal of such a situation lies,
paradoxically, in the possibility of a {\em loss} of the reality of
the spectrum. This happens, naturally, at the boundary
$\partial{\cal D}$, i.e., at a certain value of parameter(s) called,
according to Kato \cite{Kato}, exceptional point(s) (EP). They
specify, in effect, the strength of a critical non-Hermiticity at
which  the {\em ad hoc} inner product ceases to exist (cf.
\cite{ali} and \cite{SIGMA})

One of the weak points of the recipe (which the authors of Ref.
\cite{Geyer} could have called ``quasi-Hermitian quantum mechanics",
QHQM) lies in the ambiguity of the definition of the {\em ad hoc}
inner product mediated by the so called metric operator
$\Theta=\Theta^\dagger>0$,
 \be
 \br \phi|\psi\kt^{(S)}:=\br \phi|\Theta|\psi\kt^{(F)}\,.
 \ee
A way has been found in the so called ${\cal PT}-$symmetric quantum
mechanics (PTSQM) as reviewed in Ref.~\cite{Carl}. The essence of
PTSQM lies in the introduction of an additional postulate
 $
 {\cal PT} H = H {\cal PT} $ (called
${\cal PT}-$symmetry of the Hamiltonian for a suitable operator  $
 {\cal PT}$) and in its combination
with another, {\em ad hoc} requirement of factorization
$\Theta^{(PT)}={\cal PC}$ where  operator ${\cal C}$ is a charge
(cf. Ref.~\cite{Carl} for more details). The nontriviality of the
{\em additional} requirement of ${\cal PT}-$symmetry proved more
than compensated by its heuristic efficiency, with ${\cal P}$
chosen, most often, as the operator of parity.

A disappointing failure of the extension of the PTSQM formalism to
the dynamical regime of scattering \cite{Jones} has been discussed
in Ref.~\cite{fund}. The obstacle has been found removable via a
return to the QHQM recipe, with the emphasis shifted from the charge
in $\Theta^{(PT)}$ to an alternative treatment of the ambiguity of
the metric.

The perceivable increase of the technical difficulty of the required
explicit construction of the metrics $\Theta\neq \Theta^{(PT)}$ has
been softened by the discretization of the coordinates, $x \to x_n$
(cf. also \cite{smear}). The common differential toy-model
Hamiltonians $H=-\triangle+V(\vec{x})$ were replaced  by their
difference-operator analogues. Typically, the role of the kinetic
energy  $-\triangle$ was taken by a tridiagonal matrix (i.e., say,
by the well known Runge-Kutta approximation of the Laplacean).
Simultaneously, the diagonal potential-energy matrix $V(x_n)$ was
generalized into a tridiagonal matrix which was not necessarily
Hermitian in ${\cal H}^{(friendly)}$.

The resulting $N$ by $N$ Hamiltonians of the generic form
 \be
 \hat{H}^{(N)}=
  \left[ \begin {array}{cccccc}
   a_1&c_1&0&\ldots&0&0
  \\
  b_2
 &a_2&c_2&0&\ldots&0
 \\0&b_3&a_3&c_3&\ddots&\vdots
 \\ \vdots&\ddots
 &\ddots&\ddots&\ddots&0
 \\ 0&\ldots&0&b_{N-1}&a_{N-1}&c_{N-1}
 \\ 0&0&\ldots&0&b_{N}&a_{N}\\
 \end {array} \right]\,
 \label{kitielor}
 \ee
may be interpreted as representing an open-end $N$-site
nearest-neighbor-interaction quantum lattice \cite{Joglekar}.

\begin{figure}[htb]                     
\begin{center}                         
\epsfig{file=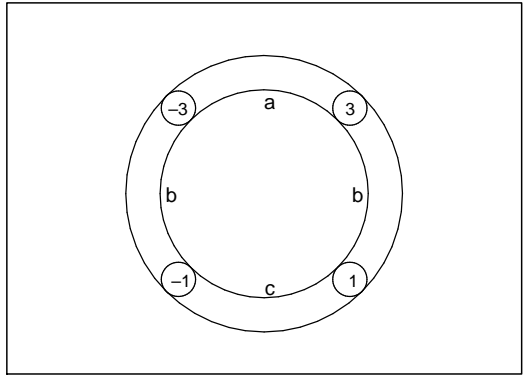,angle=270,width=0.5\textwidth}
\end{center}                         
\vspace{-2mm}\caption{A graphical representation of a sample $N=4$
lattice (\ref{kitiel}) with $a_4=-a_1=3$ and $a_3=-a_2=1$ while
$a=c_4$, $b=c_3=c_1$ and $c=c_2$.
 \label{fi3}}
\end{figure}

\begin{figure}[h]                     
\begin{center}                         
\epsfig{file=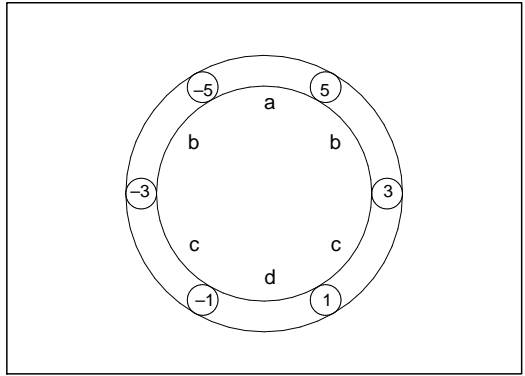,angle=270,width=0.5\textwidth}
\end{center}                         
\vspace{-2mm}\caption{A graphical representation of a sample $N=6$
lattice (\ref{kitiel}) with $a_6=-a_1=5$, $a_5=-a_2=3$ and
$a_4=-a_3=1$ while $c_6=a$, $c_5=c_1=b$, $c_4=c_2=c$ and $c_3=d$.
 \label{sefi}}
\end{figure}

In what follows we intend to complement the existing studies of
various open-end versions of the $N$-site quantum
lattice~(\ref{kitielor}) (cf., e.g., Refs.~\cite{chain} or
\cite{laguerre}, with further references) by the next-step study of
its generalization
 \be
 \hat{H}^{(N)}=
  \left[ \begin {array}{cccccc}
   a_1&c_1&0&\ldots&0&-c_N
  \\
  -c_1
 &a_2&c_2&0&\ldots&0
 \\0&-c_2&a_3&c_3&\ddots&\vdots
 \\ \vdots&\ddots
 &\ddots&\ddots&\ddots&0
 \\ 0&\ldots&0&-c_{N-2}&a_{N-1}&c_{N-1}
 \\ c_N&0&\ldots&0&-c_{N-1}&a_{N}\\
 \end {array} \right]\,,\ \ \ \ \ \ N = 2J\,.
 \label{kitiel}
 \ee
This Hamiltonian matrix may be read as representing a circular,
ring-shaped discrete lattice (cf. its graphical representation in
Figs.~\ref{fi3} and ~\ref{sefi}).

The motivation of such a project is twofold. Firstly, the
introduction of the ``anomalous" matrix elements $c_N$  enables us
to study the spectral consequences of the tunable input interaction
of a long-range character. Section \ref{pertur} will pay particular
attention to the effects of this long-range perturbation on a {\em
maximally degenerate} EP singularity at $N=6$. We shall sample
alternative scenarios of the removal of this degeneracy.

Secondly, we feel motivated by the simplicity-preserving character
of the transition from Eq.~(\ref{kitielor}) to Eq.~(\ref{kitiel}).
In section  \ref{nepertur} a few exactly solvable benchmark models
will be presented, demonstrating an enrichment of the variability of
the spectrum in non-perturbative regime. Our models will exhibit
{\em multiple complexifications} of the energies at the EP boundary
$\partial{\cal D}$.

Our observations and proposals will be finally summarized in section
\ref{VIs}.

\section{The decays of a multiple degeneracy
\label{pertur}}

\begin{figure}[h]                     
\begin{center}                         
\epsfig{file=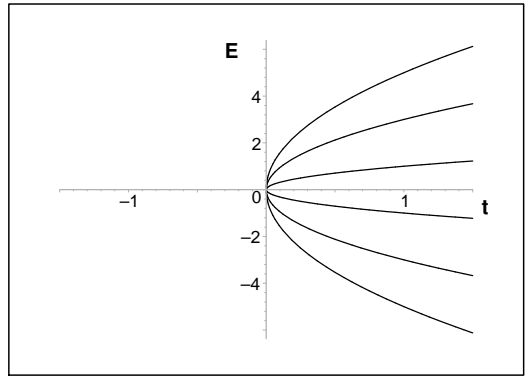,angle=270,width=0.5\textwidth}
\end{center}                         
\vspace{-2mm}\caption{The onset of the reality of the spectrum of
energies of the six-site lattice~(\ref{sixsit}) at the Kato's
degenerate exceptional point $t=0$ (the open-end dynamical regime).
 \label{fiebb}}
\end{figure}

In Ref.~\cite{chain} we proposed a family of multi-parametric
$N-$dimensional matrices (\ref{kitielor}) for which one is able to
construct certain parts of the boundary $\partial{\cal D}$ in closed
form, non-numerically. In an $N=6$ illustration of such a toy-model
scenario let us recall the multiple-degeneracy-generating
Hamiltonian matrix $H_{(MDG)}=$
 \be
  =\left[ \begin {array}{cccccc} -5&\sqrt {5-5\,t}&0&0&0&0\\
 \noalign{\medskip}-\sqrt {5-5\,t}&-3&2\,\sqrt {2-2\,t}&0&0&0
 \\\noalign{\medskip}0&-2\,\sqrt {2-2\,t}&-1&3\,\sqrt {1-t}&0&0
 \\\noalign{\medskip}0&0&-3\,\sqrt {1-t}&1&2\,\sqrt {2-2\,t}&0
 \\\noalign{\medskip}0&0&0&-2\,\sqrt {2-2\,t}&3&\sqrt {5-5\,t}
 \\\noalign{\medskip}0&0&0&0&-\sqrt {5-5\,t}&5\end {array} \right].
 \label{sixsit}
 \ee
The corresponding spectrum of energies stays unobservable (complex)
at the negative ``times" $t<0$ while becoming, suddenly, completely
degenerate at $t=0$ and strictly real and fully non-degenerate at
all the positive $t>0$. This fine-tuned multiple-degeneracy behavior
of the spectrum (sampled here in Fig.~\ref{fiebb}) may immediately
be extended to any dimension $N$ \cite{chaindva}.

We shall check what may happen when the linear chain of
Eq.~(\ref{kitielor}) is being replaced by the ring-shaped chain
represented by Eq.~(\ref{kitiel}).

\subsection{${\cal PT}-$symmetry in lattices  \label{Vs}}

It has recently been hinted in the literature that the open-end
non-Hermitian quantum lattices are exceptional ``since periodic
boundary conditions are incompatible with the  ${\cal PT}-$symmetry"
\cite{Joglekar}. For several reasons, such a comment attracted our
attention. First of all, we felt puzzled by the potential physical
consequences of such a statement since it contradicted our older
empirical experience with the existence of strong parallels between
the open- and coupled-end quantum lattices \cite{aindva}.

During our subsequent analysis of the problem we imagined that its
very core is entirely artificial. The source of the misunderstanding
may simply be traced back to certain ambiguity of the current
terminology. In an introductory remark let us, therefore, recall a
few relevant definitions while restricting our present attention,
for the sake of simplicity, just to Hamiltonians (\ref{kitielor})
or~(\ref{kitiel}) with {\em real} matrix elements.

In the first step let us slightly simplify the complex-Hamiltonian
conventions as accepted in Ref.~\cite{Joglekar}. Thus, our present
interpretation of the ${\cal PT}-$symmetry property $H{\cal
PT}={\cal PT}H$ of the real Hamiltonians will employ the
time-reversal ${\cal T}$ represented just by the operator of matrix
transposition. Secondly, the specific choice of the parity operator
 \be
 {\cal P}=
 \left[ \begin {array}{cccccc}
   1&0&\ldots&&\ldots&0  \\
  0 &-1&0&\ldots&\ldots&0 \\
 0&0&1&0&\ldots&0 \\
 0&0&0&-1&\ddots&\vdots \\
  \vdots&\ddots &\ddots&\ddots&\ddots&0 \\
  0&\ldots&0&0&0&\mp 1
   \end {array} \right]\,
   \label{parita}
 \ee
will be assumed fixed in advance. Thirdly, let us clearly
distinguish between the concepts of the so called unbroken and
broken ${\cal PT}-$symmetry where, by definition, the whole spectrum
of $H$ is real or not real, respectively.

In a more explicit concise explanation involving, for the sake of
brevity, just the non-Hermitian ${\cal PT}-$symmetric Hamiltonians
(i.e., matrices $H\neq H^\dagger$ such that $H^\dagger{\cal P} =
{\cal P}H$) with non-degenerate (though, in general, complex)
spectra, one must distinguish between the right-eigenvector
Schr\"{o}dinger equation $H|R_n\kt = E_n|R_n\kt$ and the
left-eigenvector Schr\"{o}dinger equation $\br L_n|H = \br
L_n|\,E_n$ (or, equivalently, $K\, |L_n\kt = F_n |L_n\kt$ where $K
\equiv H^\dagger \neq H$  and $F_n\equiv E_n^*$).

In this notation one easily separates the real-energy case ``A" (in
which $F_n=E_n$) and the complex-energy case ``B" (in which
$F_n=E_n^* = E_m$ at a subscript $m =m(n)\neq n$). As long as we can
always write ${\cal P}H|R_n\kt = H^\dagger \left ({\cal
P}|R_n\kt\right ) = E_n \left ({\cal P}|R_n\kt\right )$ in both of
these cases, it is easy to conclude that
 \be
 {\cal P}|R_n\kt = const\, |L_n\kt \ \ \ {\rm iff}\ \ \ E_n=E_n^*\,.
 \label{symwf}
 \ee
In the major part of the current literature on ${\cal PT}-$symmetric
Hamiltonians, the validity of proportionality (\ref{symwf}) between
the left and right eigenvectors {\em at all} $n=1,2,\ldots,N$ is,
conveniently, called the unbroken ${\cal PT}-$symmetry of the
quantum system in question. Thus, one must be a bit careful when
reading  Ref.~\cite{Joglekar} where the models with unbroken ${\cal
PT}-$symmetry are called ``models in ${\cal PT}-$symmetric phase".

This being explained, we believe that there is no true reason for
taking the circular lattices (i.e., in our case, systems with
Hamiltonians (\ref{kitiel}) mimicking the periodic boundary
conditions and representing the circular lattices of the shape
sampled in Fig.~\ref{sefi}) as a perceivably more complicated
option. We might even conjecture that as long as the circular shape
of the lattice may be perceived as an elementary exemplification of
a topologically nontrivial quantum graph of a non-tree shape, the
presence of the end-point bonds might be interpreted, in the spirit
of Ref.~\cite{anomal}, as a hidden source of potentially interesting
anomalies in the spectrum.

We shall restrict our attention to the even$-N$ subset of models
(\ref{kitiel}). The main reason is that under this restriction our
Hamiltonians will exhibit more parallels with their
differential-equation predecessors. In particular, we shall always
employ just the manifestly coupling-independent operator ${\cal P}$
of Eq.~(\ref{parita}) which strongly resembles the standard parity
with its equal number of positive and negative eigenvalues.

\subsection{A destabilization via a coupling between ends\label{prece}}

\begin{figure}[h]                     
\begin{center}                         
\epsfig{file=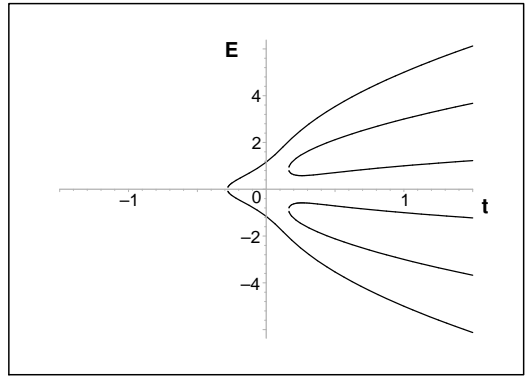,angle=270,width=0.5\textwidth}
\end{center}                         
\vspace{-2mm}\caption{The change of the spectrum of Fig.~\ref{fiebb}
after the transition to the ring-shaped lattice of Fig.~\ref{sefi}
in the weak-coupling dynamical regime.
 \label{fiedeeb}}
\end{figure}

 \noindent
For a sampling of the effects of the periodicity-simulating
lattice-ends couplings  $\pm c_N$ let us return to the six-site
open-end-lattice spectrum of Fig.~\ref{fiebb} and let us treat the
bonding matrix elements $\pm c_N$ as a small perturbation to this
unperturbed form of the ${\cal PT}-$symmetric Hamiltonian.

Let us consider the first sample of such a perturbed Hamiltonian
$H^{(6)}_1(t)=H_{(MDG)}+W_1$, with the perturbation specified as
follows,
 \ben
 W_1= \left[ \begin {array}{cccccc} 0&0&0&0
  &0&-w\\
 \noalign{\medskip}0&0&0&0&0&0
 \\\noalign{\medskip}0&0&0&0&0&0
 \\\noalign{\medskip}0&0&0&0&0&0
 \\\noalign{\medskip}0&0&0&0&0&0
 \\\noalign{\medskip}w
 &0&0&0&0&0\end {array}
 \right]\,,\ \ \ w =\frac{1}{100}\sqrt {1-t}\,.
 \een
The resulting modified $t-$dependence of the spectrum is displayed
in Fig.~\ref{fiedeeb}. We see that the original collapse of the
whole spectrum  gets split. A twin partial collapse is shifted to
the right (i.e., to $t\to 0. 159^+$) while the ultimate complete
complexification moved to the left (i.e., to negative
$t\to-0.2818^+$).

\subsection{Competing agents of destabilization}

\begin{figure}[h]                     
\begin{center}                         
\epsfig{file=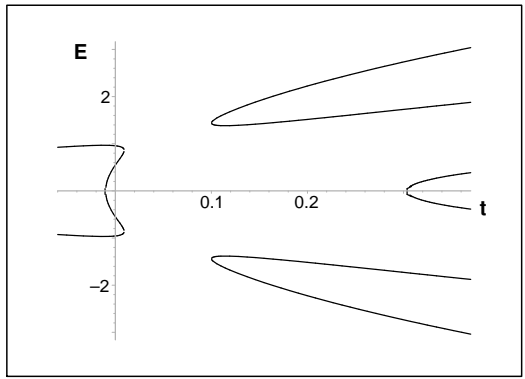,angle=270,width=0.5\textwidth}
\end{center}                         
\vspace{-2mm}\caption{The energies of the six-site
lattice~(\ref{fifi}) (Fig.~\ref{sefi}) with a stronger bond between
the endpoints and with an enhanced central attraction.
 \label{fiedeece}}
\end{figure}

\begin{figure}[h]                     
\begin{center}                         
\epsfig{file=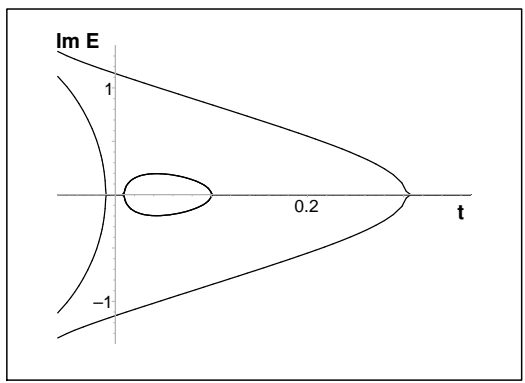,angle=270,width=0.5\textwidth}
\end{center}                         
\vspace{-2mm}\caption{A pendant of Fig.~\ref{fiedeece} -- the
imaginary parts of the energies.
 \label{fiedeeef}}
\end{figure}

 \noindent
From the observations made in paragraph \ref{prece} one can conclude
that both the maximal degeneracy (at $t=0$) and the Big-Bang-like
behavior (at the small $t>0$)  of the spectrum as sampled by
Fig.~\ref{fiebb} are in fact very sensitive to perturbations. The
form of this perturbation is highly relevant for the guarantee of
the stability of the perturbed quantum system, i.e., of the reality
of its bound-state energies.

The determination of the boundary $\partial{\cal D}$ of the domain
${\cal D}$ of the parameters which are compatible with the stability
is important. Certain qualitative features of this boundary (or
``horizon") may even be guessed {\em a priori}. Typically, one may
expect that the system gets less stable, i.e., the size of the
domain ${\cal D}$ will shrink in the strong-coupling regime. In
parallel, the destabilizing effect of the growth of the end-bond
$c_6=a$ (cf. Fig.~\ref{sefi}) may be compensated by the decrease
(or, alternatively, enhanced by the further increase) of some other
non-Hermiticity measure (i.e., for the sake of definiteness, of
$c_3=d$ in Fig.~\ref{sefi}).

The consequences are both interesting and hardly predictable.
Figures~\ref{fiedeece} and \ref{fiedeeef} offer an illustration of
the effect. These pictures illustrate a scenario of destabilization
in which, in spite of the perceivable growth of $a$ (i.e., in spite
of our making the ring better bound), the dominant role is still
played by the smaller perturbation of the close-to-critical $d$.

This guess may be assisted by the toy-model Hamiltonian
$H^{(6)}_2(t)=H_{(MDG)}+W_2$ with
 \be
 W_2= \left[ \begin {array}{cccccc} 0&0&0&0
  &0&-10w\\
 \noalign{\medskip}0&0&0&0&0&0
 \\\noalign{\medskip}0&0&0&w&0&0
 \\\noalign{\medskip}0&0&-w&0&0&0
 \\\noalign{\medskip}0&0&0&0&0&0
 \\\noalign{\medskip}10w
 &0&0&0&0&0\end {array}
 \right]\,,\ \ \ w =\frac{1}{100}\sqrt {1-t}\,.
 \label{fifi}
 \ee
The mechanism of the dominance of the enhancement of the central
coupling is demonstrated by Figs.~\ref{fiedeece} and \ref{fiedeeef}.
We see there the remarkable pattern of complexification in which the
rightmost exceptional point $t_{+} \approx 0.3$ of the loss of the
crypto-Hermiticity (i.e., of the reality of the whole spectrum) is
determined by $d$ while the next coordinate $t_{-} \approx 0.1$ of
the remaining two mergers and subsequent complexifications already
reflects the combined effect of the whole perturbation.

A qualitative novelty may be seen in the re-emergence of a small
island of the reality of as many as four central energies in a  very
small vicinity of  $t=0$. This phenomenon is a close analogue of the
similar spectral ``reality-island" anomaly encountered, in
Ref.~\cite{anomal}, in the strong-coupling regime of another
topologically nontrivial model.

\section{Decays of separate degeneracies \label{nepertur}}


The explicit evaluation of the periodic-lattice spectra remains a
more or less purely numerical problem in the tight-binding regime,
especially at the larger $N$. One still encounters exactly solvable
secular equations at the smallest even $N=2J$. {\it In extremis},
many generic features of the periodic-lattice spectra may be
understood even via their first nontrivial four-site-lattice
realization.

\subsection{An exactly solvable model}

\begin{figure}[h]                     
\begin{center}                         
\epsfig{file=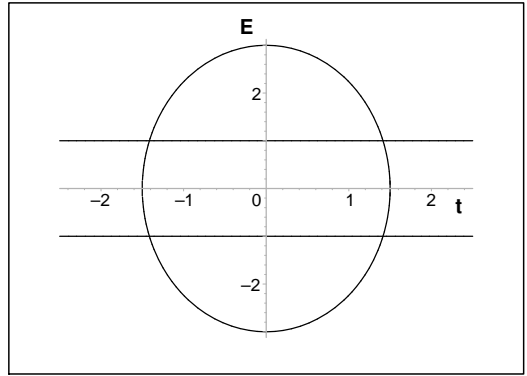,angle=270,width=0.5\textwidth}
\end{center}                         
\vspace{-2mm}\caption{The spectrum of energies of the periodic
equal-coupling four-site-lattice Hamiltonian (\ref{away}).
 \label{fiedeero}}
\end{figure}

 \noindent
Once we pick up the following one-parametric four-site toy model
with equal couplings,
 \be
 H_{(EC)}=\left[ \begin{array}{cccc}
 -3&t&0&-t\\\noalign{\medskip}-t&-1&t&0
 \\\noalign{\medskip}0&-t&1&t
 \\\noalign{\medskip}t&0&-t&3\end{array}
 \right]
 \label{away}
 \ee
we reveal that the whole spectrum remains real up to the
strong-coupling dynamical regime, i.e., even beyond $|t|=1$ (cf. the
graphical representation of this spectrum in Fig.~\ref{fiedeero}).
For our present purposes it is also useful that the spectrum is in
fact available in closed form,
 \be
 E_0=- \sqrt{9-4\,t^2}\,,\ \ \ \
 E_1=-1\,,\ \ \ \
 E_2=1\,,\ \ \ \
 E_3=\sqrt{9-4\,t^2}\,.
 \label{twoe}
 \ee
One localizes, precisely,  the unavoided-crossing points
$t=t_{\pm}^{(UC)} = \pm \sqrt{2}$ as well as the points of the
ultimate complexification $t=t_{\pm}^{(C)} = \pm {3/2}$. All of
these four values  are the exceptional points in the Kato's sense
\cite{Kato}. In order to see this, one may recall Eq.~(\ref{twoe})
and pick up, say, eigenvalues $E_2$ and $E_3$ of our Hamiltonian
(\ref{away}). Then, the two respective eigenvectors, viz., the
four-component vector $\vec{\psi}_2=[0, t, 2, t]$ and its partner
$\vec{\psi}_3$ with components
 \ben
 \left [{t}^{2}-2,\,
  \left( \sqrt {9-4\,{t}^{2}}+1 \right)\,t/2 ,3-{t}^{2}+
 \sqrt {9-4\,{t}^{2}},\left [\vec{\psi}_3\right ]_4 \right ]\,,
 \een
 \ben
 \left [\vec{\psi}_3\right ]_4={\frac {(4-{t}^{2})
 \,\sqrt {9-4\,{t}^{2}}+12-5\,{t}^{2}}{2\,t}}
 \een
will strictly coincide in either of the limits of $t \to
t_{\pm}^{(UC)}$. The Hamiltonian will only remain diagonalizable
(i.e., crypto-Hermitian) inside the three separate intervals of $t$,
viz, inside domain ${\cal D}^{(H)}_0= (-\sqrt{2},\sqrt{2})$
(combining the weak- and strong-coupling dynamical regimes) or
inside ${\cal D}^{(H)}_-= (-3/2,-\sqrt{2})$ or ${\cal D}^{(H)}_+=
(\sqrt{2},3/2)$ (= the two perceivably smaller domains of
anomalously strong couplings).

Inside the non-anomalous domain ${\cal D} = (-\sqrt{2},\sqrt{2})$ of
$t$ our choice of the first nontrivial $N=4$ enables us  to find a
{\em complete} family of all of the candidates for the metric in
principle. A sample of the necessary linear algebra reconstructing,
basically, the metric from the relation
 \be
 H^\dagger=
 \Theta\,H\,\Theta^{-1}\,
 \label{dieu}
  \ee
may be found, say, in Ref.~\cite{fund}. An entirely exhaustive
explicit construction of the metrics $\Theta$ has been performed
there for a certain extremely elementary one-parametric model. Here,
we shall only pick up a single, characteristic solution
 \be
 \Theta=\left[ \begin {array}{cccc} 3+{t}^{2}&-3\,t&{t}^{2}&t
 \\\noalign{\medskip}-3\,t&3+{t}^{2}&-3\,t&{t}^{2}\\\noalign{\medskip}{
t}^{2}&-3\,t&3+{t}^{2}&-3\,t\\\noalign{\medskip}t&{t}^{2}&-3\,t&3+{t}^
{2}\end {array} \right]\, \label{metra2}
 \ee
as a candidate for the metric in ${\cal H}^{(sophisticated)}$. In
order to confirm its eligibility we must prove that it is positive
definite. Such a proof is easy since the two eigenvalues of our
candidate matrix happen to possess the elementary form
 $$
3+{t}^{2}+t \pm \sqrt {13\,{t}^{2}+{t}^{4}+6\,{t}^{3}}\,.
 $$
The other two eigenvalues are obtained by the replacement $t \to
-t$. This implies the positivity of our matrix (\ref{metra2}) inside
the interval of
 \ben
 t \in (-\sqrt{3/2},\sqrt{3/2}):={\cal D}_\Theta \approx
(-1.225,1.225).
 \een
We see that our choice of special metric is satisfactory since this
interval covers more than 86 \% of the whole crypto-Hermiticity
domain of the Hamiltonian itself.


\subsection{The strengthened bond }

\begin{figure}[h]                     
\begin{center}                         
\epsfig{file=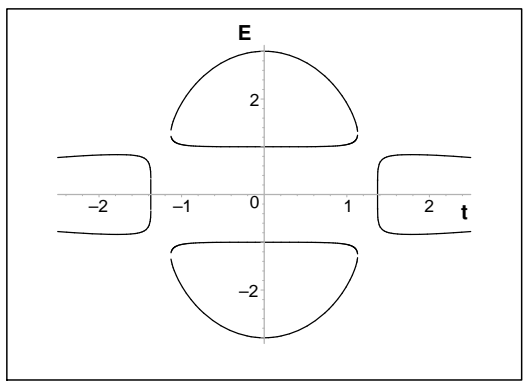,angle=270,width=0.5\textwidth}
\end{center}                         
\vspace{-2mm}\caption{The change of the spectrum of
Fig.~\ref{fiedeero} caused by the 50\% enhancement of the value of
the periodicity-guaranteeing bond.
 \label{fiedeese}}
\end{figure}

Let us now replace the fine-tuned one-parametric four-site  model
(\ref{away}) by its perturbation
 \be
 H=
 H_{(EC)}+
 \left[ \begin{array}{cccc}
 0&0&0&-t/2
 \\\noalign{\medskip}0&0&0&0
 \\\noalign{\medskip}0&0&0&0
 \\\noalign{\medskip}t/2&0&0&0
 \end{array}
 \right]\,.
 \label{awayab}
 \ee
Figure \ref{fiedeese} illustrates the changes. The decay of the two
original point-like unavoided crossings at $t_{\pm}^{(UC)}$ spreads
over the respective two small non-empty intervals of $t$ in which
all of the four eigenenergies become complex, acquiring a
non-vanishing imaginary part.

A more detailed inspection of Fig.~\ref{fiedeese} reveals that the
central domain shrinks a bit in comparison, ${\cal D}^{(H)}_0
\approx (-1.13137,1.13137)$. The ultimate onset of the large$-t$
survival of the two real energies moves from the points
$t_{\pm}^{(C)}=3/2$ slightly closer to the origin.

\begin{figure}[h]                     
\begin{center}                         
\epsfig{file=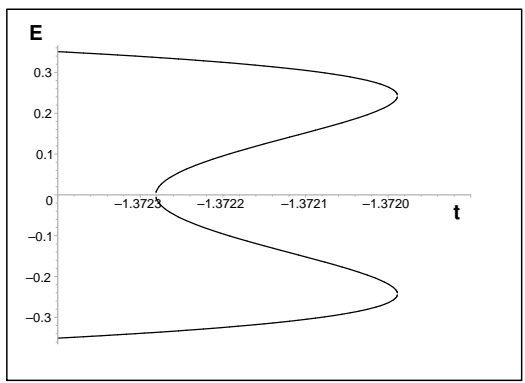,angle=270,width=0.5\textwidth}
\end{center}                         
\vspace{-2mm}\caption{Same as Fig.~\ref{fiedeese}, in a more
detailed blow up.
 \label{fiedeesese}}
\end{figure}

Figure~\ref{fiedeese} seems to indicate the complete disappearance
of the separate anomalous domains ${\cal D}^{(H)}_\pm$. The
magnified version of the same picture (cf. Fig.~\ref{fiedeesese})
reveals that it is not so. The two very small but still non-empty
intervals ${\cal D}^{(H)}_+ \approx (1.37199,1.37228)$ and ${\cal
D}^{(H)}_- \approx (-1.37228,-1.37199)$ survive and yield the whole
quadruplet of the energies purely real again, in a truly fine-tuned
manner.

A challenging task emerges in connection with an appropriate
modification of the metric. In place of Eq.~(\ref{metra2}) the
present would-be positive-definite solution of Eq.~(\ref{dieu})
reads
 $$
\left[ \begin {array}{cccc} 3+{t}^{2}&{\frac { \left( 3+{t}^{2}
\right) t \left( 13\,{t}^{2}-96 \right) }{17\,{t}^{2}+96}}&24\,{
\frac { \left( 3+{t}^{2} \right) {t}^{2}}{17\,{t}^{2}+96}}&1/2\,{
\frac { \left( 3+{t}^{2} \right) t \left( {t}^{2}+96 \right)
}{17\,{t} ^{2}+96}}\\\noalign{\medskip}{\frac { \left( 3+{t}^{2}
\right) t
 \left( 13\,{t}^{2}-96 \right) }{17\,{t}^{2}+96}}&3+{t}^{2}&{\frac {
 \left( 3+{t}^{2} \right) t \left( 7\,{t}^{2}-96 \right) }{17\,{t}^{2}
+96}}&24\,{\frac { \left( 3+{t}^{2} \right)
{t}^{2}}{17\,{t}^{2}+96}}
\\\noalign{\medskip}24\,{\frac { \left( 3+{t}^{2} \right) {t}^{2}}{17
\,{t}^{2}+96}}&{\frac { \left( 3+{t}^{2} \right) t \left(
7\,{t}^{2}- 96 \right) }{17\,{t}^{2}+96}}&3+{t}^{2}&{\frac { \left(
3+{t}^{2}
 \right) t \left( 13\,{t}^{2}-96 \right) }{17\,{t}^{2}+96}}
\\\noalign{\medskip}1/2\,{\frac { \left( 3+{t}^{2} \right) t \left( {t
}^{2}+96 \right) }{17\,{t}^{2}+96}}&24\,{\frac { \left( 3+{t}^{2}
 \right) {t}^{2}}{17\,{t}^{2}+96}}&{\frac { \left( 3+{t}^{2} \right) t
 \left( 13\,{t}^{2}-96 \right) }{17\,{t}^{2}+96}}&3+{t}^{2}
\end {array} \right]\,.
 $$
In spite of its perceivably more complicated structure this matrix
may still be shown, by the same techniques as above, to be safely
positive definite (i.e., to become eligible as a metric) for
 \ben
 t \in {\cal D}_\Theta \approx
 (-1.082854389,1.082854389)\,
  \een
i.e., in a slightly diminished range of the ``time" parameter.

\subsection{A recoupled regime}

In a way complementing Eq.~(\ref{awayab}) let us consider
 \be
H=
 H_{(EC)}+\left[ \begin{array}{cccc}
 0&0&0&3t/4
 \\\noalign{\medskip}0&0&t/3&0
 \\\noalign{\medskip}0&-t/3&0&0
 \\\noalign{\medskip}-3t/4&0&0&0
 \end{array}
 \right]\,.
 \label{awayaber}
 \ee
The fine-tuned model (\ref{away}) looks now perturbed in an opposite
direction. The perceivable weakening of the matrix element in the
corner (i.e., of the periodicity-guaranteeing bond) enhances the
parallels with the open-end systems. By itself, this change should
lead to a completely complex spectrum at the larger $|t|$s.

\begin{figure}[h]                     
\begin{center}                         
\epsfig{file=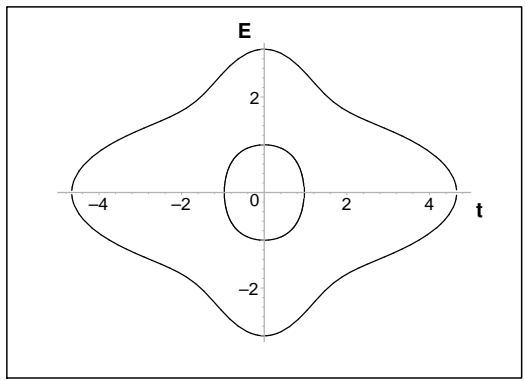,angle=270,width=0.5\textwidth}
\end{center}                         
\vspace{-2mm}\caption{The thorough deformation of the spectrum of
Fig.~\ref{fiedeero} caused by the recompensation plus weakening of
the periodicity-guaranteeing bond.
 \label{fiedeeze}}
\end{figure}

This expectation is confirmed by Fig.~\ref{fiedeeze}. The picture
shows that the choice of the stronger central attraction in
Eq.~(\ref{awayaber}) works in the fragility-enhancing direction,
diminishing the central crypto-Hermiticity domain,
 \ben
 {\cal D}^{(H)}_0:=
 \left (
 {\frac {3\,\sqrt {97}-45}{16}},
 {\frac {45-3\,\sqrt {97}}{16}}
 \right )
 \approx
 (-0.9658391622, 0.9658391622)\,.
 \een
The presence of the inflexions in the outer energy loop finds its
origin in the highly unstable unavoided-crossing points of model
(\ref{away}). The relevance of the inflexion points as emphasized in
Ref.~\cite{Joglekar} might be recalled.

The construction of the metric preserves the algebraic structure
shown above. From an updated formula (using just different numerical
coefficients) we derive the boundary points of the interval ${\cal
D}_\Theta$. The definition of these boundaries is provided by the
minimal root of the expression
 \ben
 \left( -{\frac
{1235}{8}}\,{t}^{3}+631\,{t}^{2}-720\,t+1152-1/24\,\sqrt{V}\right )
\left( 3+{t}^{2} \right)
 \left( 631\,{t}^{2}+1152 \right) ^{-1}\,,
 \ben
 \ben
 V={2772145\,{t}^{6}-50595840\,{t}^{5}+348491520\,{t}^{4}-955514880
 \,{t}^{3}+1147944960\,{t}^{2}}\,
 \een
giving
 \ben
 t \in {\cal D}_\Theta \approx
 (-0.9658391622,0.9658391622)\,.
  \een
The size of this interval of the positivity of $\Theta$ is again
comparable with the size of the interval of the crypto-Hermiticity
of the Hamiltonian.

\section{Outlook \label{VIs}}

The recent growth of interest in ${\cal PT}-$symmetric quantum
lattices offers a natural motivation of the transition to the
loop-shaped lattices. We found that a core of the consistency of
such a transition (which could suffer from its potential
``fragility" in general \cite{fragile}) is  similar to the
suppression of the fragility in open-ends models.

Via a thorough description of a few not too complicated examples we
illustrated that in {\em both} the open-end and matched-ends models,
the stability and the robust nature (i.e., non-fragile nature) of
the models results from the absence of the degeneracy of spectra of
the zero-coupling versions of the Hamiltonians in question.

The role of the matching matrix elements of the Hamiltonian remains,
in the phenomenological perspective, slightly counterintuitive.
Still, the decisive conceptual parallels between periodic lattices
and the mathematically more friendly open-end lattices were noticed.
They involve not only the encouraging emergence of the parallel
structures in the spectra of energies but also in the shapes of the
domains of the reality of the energies or of the positivity of the
matrix candidates for the metrics.

An important merit of our specific models may be seen in the
feasibility of calculations. This resulted from the preservation of
multiple parallels between our present matched-ends models and their
open-ends predecessors (let us mention just the up-down symmetry,
equidistance of the unperturbed spectra or the reality of the
interaction matrices). Nevertheless, even beyond the level of the
low-dimensional solvable examples the more general questions of
consistency of the underlying quantum theory were addressed. Our
constructive study of the  chains defined along discrete loops
appeared more friendly than expected.

Our detailed analyses covered the extensive dynamical domain, far
beyond the mere weak-coupling subdomain. Our  periodicity-simulating
bonds proved connected to the emergence of unexpected spectral
phenomena (like the strong-coupling-related islands of stability)
which will certainly deserve a further study in the future.

\subsection*{Acknowledgments}

Work supported by the GA\v{C}R grant Nr. P203/11/1433, by the
Institutional Research Plan AV0Z10480505 and by the M\v{S}MT
``Doppler Institute" project Nr. LC06002.

\end{document}